\documentclass[english,aps,prl,twocolumn]{revtex4-1}
\usepackage[T1]{fontenc}
\usepackage[latin9]{inputenc}
\usepackage{amstext}
\usepackage{stmaryrd}
\usepackage{babel}
\begin{document}

\title{Simulability of Imperfect Gaussian and Superposition Boson Sampling}

\author{Jelmer Renema}

\affiliation{Complex Photonic Systems Group, Mesa+ Institute for Nanotechnology,
University of Twente, PO Box 217, 7500 AE Enschede, The Netherlands}
\begin{abstract}
We study the hardness of classically simulating Gaussian boson sampling
at nonzero photon distinguishability. We find that similar to regular
boson sampling, distinguishability causes exponential attenuation
of the many-photon interference terms in Gaussian boson sampling.
Barring an open problem in the theory of matrix permanents, this leads
to an efficient classical algorithm to simulate Gaussian boson sampling
in the presence of distinguishability. We also study a new form of
boson sampling based on photon number superposition states, for which
we also show noise sensivity. The fact that such superposition boson
sampling is not simulable with out method at zero distinguishability
is the first evidence for the computational hardness of this problem. 
\end{abstract}
\maketitle
The next milestone in experimental photonic quantum information processing
is the demonstration of a quantum advantage: a well-defined computational
task at which a quantum device outperforms a classical computer \cite{Wolf2018,Preskill2018,Harrow2017,Arute2019}.
The photonic implementation of this concept is boson sampling \cite{Aaronson2011},
which consists of sending single photons through a linear optical
network followed by photodetection, a task which is strongly believed
to be hard for a classical computer to simulate. The best known algorithm
to clasically simulate a boson sampler is due to Clifford \& Clifford
\cite{Clifford2017}, and generates a sample in $m2^{m}$ steps, where
$m$ is the number of photons. Given this scaling and the speed of
modern supercomputers, it is believed that a boson sampler will outperform
a classical simulation on a supercomputer around $m\approx50$ photons
\cite{Neville2017,Wusupercomputer}. Experimentally demonstrating
boson sampling is the subject of experimental efforts worldwide \cite{Broome2012,Spring2012,Tillmann2013,Crespi2013,Carolan2015,Bentivegna2015,Wang2017,Wang2018,Wang2019b}

A major problem in boson sampling theory is understanding the degree
to which quantum photonic interference is susceptible to imperfections
\cite{Shchesnovich2014,Shchesnovich2015a,Shchesnovich2015b,ShchesnovichQMoivre,ShchesnovichSciRep,Tamma2015,Tamma2015PRL,Tichypartial,Laibacher2018}.
These imperfections can take the form of any experimental noise which
degrades the quantum nature of the observed interference pattern.
The two imperfections most strongly present in experiments are photon
loss and distinguishability. Distinguishability is the imperfect overlap
of the wavefunctions of the photons in the other degrees of freedom
besides their position (e.g. polarization, frequency, time). Loss
is when not all photons produced by the sources reach the experiment.
Other imperfections include dark counts (detection events not associated
with the arrival of a photon) and fluctuations in the settings of
the interferometer. 

In a series of recent works \cite{Kalai2014,Renema2018a,Renema2018b,Moylett2019,Shchesnovich2019a,Shchesnovich2019b},
it was shown that boson samplers with have any of these imperfections
at a sufficiently strong level can be efficiently clasically simulated,
i.e. in polynomial time, thereby negating any potential quantum advantage.
The strategy behind these simulations is to split the interference
pattern into a series of $j$-photon interference terms, where $j$
runs from 0 to the number of photons $m$. Imperfections degrade the
higher-order interference terms more strongly than the lower orders,
which means that for sufficiently strong imperfections, the series
over $j$ can be truncated at some value $k$. The resulting algorithm
is polynomial in $m$. The strength of the imperfections determines
the maximum level of interference $k,$ thereby functioning as a criterion
for simulability of $k$-photon interference. The physical picture
arising from these results is that imperfect boson sampling can be
thought of as interference between all possible groups of $k$ photons
and classical (i.e. single-particle) transmission of the remaining
$m-k$ photons. 

However, these simulability criteria do not automatically extend to
boson sampling variants. Variants on boson sampling exist as workarounds
for the fact that high efficiency, deterministic single photon sources
are not available experimentally. The two main kinds of photon sources
currently available are those based on single-photon emission from
quantum dots, where a $\pi$ Rabi flip is used to excite a two-level
system \cite{Gazzano2013,Dousse2010,Somaschi2016}, which then spontaneously
emits a photon, and those based on spontaneous parametric down-conversion
(PDC) in nonlinear optical media such as $\beta$-barium borate and
potassium titanyl phosphate \cite{Boyd2008}. 

While quantum dot sources have seen extremely rapid progress in recent years
\cite{Somaschi2016,Wang2019a}, PDC sources still hold the record
for photon collection efficiency (which is the relevant quantity for
complexity analysis) and indistinguishability. However, their main
weakness is that they do not emit single photons but two-mode squeezed
states, which are nonclassical states consisting of a thermal distribution
of pairs of photons, including zero pairs. The presence of a zero-photon
component in the state means that these sources cannot be made to
function deterministically. To circumvent the fact that these sources
are non-deterministic, the usual strategy is heralding: by detecting
one photon, the existence of the other can be inferred, and it can
then be used for experiments \cite{Clauser1974}. Such heralded nondeterministic
sources can be used to sample over a classical probability distribution
of input states by recording which sources produced a photon in any
given run of the experiment. This problem, known as scattershot boson
sampling, directly derives its hardness from the original boson sampling
problem \cite{Aaronsonweb,Lund2014}. 

A further step in this line of thought is to dispense with the heralds
altogether, and feed both photons into the linear network \cite{Hamilton2017}.
This protocol is known as Gaussian boson sampling (GBS), after the
fact that a squeezed state is a Gaussian state. For this situation,
the reduction to a known hardness argument is via the case where the
linear optical transformation effected by the interferometer is separable
into one acting on the signal modes and one acting on the herald modes
\cite{Chakhmakhchyan2017}, or equivalently where the action on the
herald modes is the identity. In that case, the problem reduces to
scattershot boson sampling \cite{Lund2014,Aaronsonweb}. It is largely
unknown how imperfections affect Gaussian boson sampling \cite{Qi2019}.

In this work, we will investigate the simulability of Gaussian Boson
Sampling (GBS) and an auxilliary model, which we call Superposition
Boson Sampling (SBS), which is inspired by recent work on emission
of coherent superposition states from quantum dots \cite{Loredo2018}.
We will consider these two systems both with and without noise. We
will show that without noise, both systems are not simulable by the
method outlined above. For superposition sampling, this is the first
evidence (although circumstantial) of computational hardness for this
system. When we introduce noise, we see that the same physical picture
arises for both systems as for regular boson sampling: the interference
pattern falls apart into smaller interference processes of size $k<m$,
where $k$ depends only on the strength of the imperfections. For
SBS, we are able to give an explicit algorithm to efficiently classically
simulate imperfect boson sampling. For GBS, we reduce the problem
of finding such an algorithm to a problem in the theory of matrix
permanents. 

We restrict ourselves to studying the effect of one particular imperfection,
namely photon distinguishability. We note, however, that for regular
(i.e. Fock-state) boson sampling, all imperfections considered in
the literature so far can be analyzed using the strategy which we
will use. This restriction is therefore not as strong as it might
at first appear. We hope to study the effect of other imperfections
on Gaussian boson sampling in later work.

The paper is organized into six sections as follows: in Section I,
we will begin by recalling a method to compute the output of a boson
sampler fed with arbitrary input states. Then in Section II, we will
re-derive the simulability criteria for regular boson sampling. Our
results begin in Section III, where we will show how to combine these
two into a criterion for simulability of boson sampling with arbitrary
input states. Next, we will apply these results to the special case
of superposition boson sampling in Section IV. Finally, we will show
in Section V that under weak pumping conditions, Gaussian boson sampling
reduces to superposition sampling. The paper ends with concluding
remarks in Section VI.

\section{boson sampling with arbitrary input states}

The theory for boson sampling with arbitrary input states was derived
in \cite{Shchesnovich2017}. Here, we provide a derivation of the
expressions for a detection probability at the output of a linear
optical network with arbitrary states at the input and finite distinguishability
for reasons of exposition, and to establish the framework used in
the present work.

Consider an arbitrary multimode photonic quantum state $|\psi\rangle$
impinging on a linear optical network $U$ with detectors in the Fock
basis at the output of the network. We wish to compute the probability
of an arbitrary pattern of detection events $|\phi_{pd}\rangle$,
and to study the hardness of computing that outcome. The probability
is given by:
\begin{equation}
P(\phi_{pd})=|\langle\psi|U|\phi_{pd}\rangle|^{2}.
\end{equation}

We can then insert a resolution of the identity in the Fock basis:
$\hat{I}_{\mathrm{fs}}=\sum_{p}|\xi_{p}\rangle\langle\xi_{p}|$, where
$|\xi_{p}\rangle=\prod_{i=1}^{n}|m_{i}\rangle$ is a product of Fock
states, and $m_{i}$ are the mode occupation numbers: 
\begin{equation}
P(\phi_{pd})=|\sum_{p}\langle\psi|\xi_{p}\rangle\langle\xi_{p}|U|\phi_{pd}\rangle|^{2}.
\end{equation}

Since the interferometer $U$ is photon-number preserving, all terms
in the sum over $\xi$ which do not contain the same number of photon
numbers as $|\phi_{pd}\rangle$ drop out, i.e. $\langle\xi_{p}|U|\phi_{pd}\rangle=0$
unless $|\langle\xi_{p}|\hat{N}|\xi_{p}\rangle|^{2}=|\langle\phi_{pd}|\hat{N}|\phi_{pd}\rangle|^{2},$
where $\hat{N}$ is the multimode photon number operator. Hence we
relabel the sum over $p$ to contain only those terms which meet this
condition. Expanding, we have:
\begin{eqnarray}
P(\phi_{pd}) & = & \sum_{p}\sum_{q}\langle\psi|\xi_{p}\rangle\langle\psi|\xi_{q}\rangle^{\dagger}\langle\xi_{p}|U|\phi_{pd}\rangle\langle\xi_{q}|U|\phi_{pd}\rangle^{\dagger}\\
 & = & \sum_{p}\sum_{q}c_{p}c_{q}^{\dagger}\mathrm{Perm}(M_{p})\mathrm{Perm(}M_{q})^{\dagger}.\nonumber 
\end{eqnarray}

Here we have used the fact that since $\xi$ is a product of Fock
states, we can apply the identity $\langle\xi_{p}|U|\phi_{pd}\rangle=\mathrm{Perm}(M_{\xi_{p}\phi})/\sqrt{\mu(\xi_{p})\mu(\phi_{pd})}$,
where Perm is the permanent function $\mathrm{Perm}(M)=\sum_{\sigma}\prod_{i}M_{i,\sigma_{i}}$,
and M is the submatrix connecting $\xi_{p}$ and $\phi_{pd}$ \cite{Scheel2008}.
Since we are concerned only with a single output $\phi_{pd},$ we
will suppress this dependence, as well as the subscript $\xi$ and
denote the matrix as $M_{p}$. The coefficients are given by $c_{p}=\langle\psi|\xi_{p}\rangle/\sqrt{\mu(\xi_{p})\mu(\phi_{pd})}.$
$\mu(\xi)$ is the multiplicity of a particular Fock state configuration:
$\mu(\xi)=\prod_{i}(m_{i}!)$. Furthermore, in what follows, we will
assume $\mu(\phi_{pd})=1,$ a condition which can be enforced with
high probability by making the linear interferometer large enough
\cite{Aaronson2011}.

Note that eqn 3 is a natural way to consider interference from a quantum
state with an indeterminate photon number in each mode: this equation
simply tallies all the ways in which the set of sources could have
produced a given number of photons, interfering with all the other
ways those sources could produce that number of photons. These interference
terms have been observed in experiments \cite{Carolan2014}.

Anticipating our complexity analysis, we reorder terms in eqn 3:
\begin{equation}
P(\phi_{pd})=\sum_{p}\sum_{q}c_{p}c_{q}^{\dagger}\sum_{\sigma}\mathrm{Perm}(M_{p}\circ M_{\sigma(q)}^{\dagger}),
\end{equation}
where $\sigma(q)$ denotes a permutation of the matrix elements of
$M_{q},$ and $\circ$ denotes the elementwise product. 

Next, we look for cases where the combination of $\xi_{p},$ $\xi_{q}$
and $\sigma$ causes a product between the $i$-th row of $M$ and
its complex conjugate to appear in the permanent. Such a row of modulus-squared
terms implies classical interference of the corresponding photon.
We will later see that these terms also correspond to the fixed points
$j$ of the partial permutation $\xi_{p}\rightarrow\sigma(\xi_{q}).$
We can split off these positive rows using Laplace expansion along
those rows:

\begin{eqnarray}
P(\phi_{pd}) & = & \sum_{p}\sum_{q}c_{p}c_{q}^{\dagger}\sum_{j}^{m}\sum_{\sigma^{j}}\mathrm{Perm}(M_{p}\circ M_{\sigma(q)})\\
= &  & \sum_{p}\sum_{q}c_{p}c_{q}^{\dagger}\sum_{j}^{m}\sum_{\sigma^{j}}\sum_{\rho}\mathrm{Perm}(M_{p,\rho}\circ M_{\sigma_{p}(q),\rho}^{\dagger})\nonumber \\
 &  & \ \ \ \ \ \ \ \ \ \ \ \ \ \ \ \ \mathrm{\ \ \ \ \ \ \ \ \ \ \ \ \ \times Perm}(|M_{p,\bar{\rho}}|^{2}),\nonumber 
\end{eqnarray}
where $\sigma_{p}$ stands for the part of $\sigma$ not corresponding
to fixed points, $\rho$ is a $j$-partition of the number of detected
photons $m,$ the sum runs over all $j$-partitions, and the overbar
denotes the complement of $\rho$ on the set $\xi_{p}$. 

We now determine the effect of distinguishability on this optical
system. From eqn 4, it is clear that for fixed $p$ and $q$, we are
considering all the ways in which photons from configurations $\xi_{p}$
and $\xi_{q}$ can give rise to a given detection event. Therefore,
when considering distinguishability, we must consider the overlap
of the internal degrees of freedom at the output side of a double-sided
Feynman diagram \cite{Tichypartial}. The prefactor to adjust for
distinguishability is given by $\prod_{i}\langle\bar{\psi}_{p_{i}}|\bar{\psi}_{\sigma_{i}(q)}\rangle,$
where $|\bar{\psi}_{q_{i}}\rangle$ denotes the internal state (degrees
of freedom unaffected by $U$) for the $i$-th photon in $q$, and
$\sigma_{i}(q)$ denotes the $i$-th element of the permutation $\sigma$
acting on $q.$ Hence we obtain:
\begin{eqnarray}
P(\phi_{pd}) & = & \sum_{p}\sum_{q}c_{p}c_{q}^{\dagger}\sum_{j}\sum_{\sigma^{j}}\left(\prod_{i}\langle\bar{\psi}_{p_{i}}|\bar{\psi}_{\sigma_{i}(q)}\rangle\right)...\\
 &  & ...\sum_{\rho}\mathrm{Perm}(M_{p,\rho}\circ M_{\sigma_{p}(q),\rho})\mathrm{Perm}(|M_{p,\bar{\rho}}|^{2}).\nonumber 
\end{eqnarray}

To simplify the analysis without losing any of the essential features,
we will assume throughout this paper that all distinguishability inner
products between the $p$-th and $q$-th photon are equal to a constant
value $x$ for $p\neq q$. In that case, the distinguishability factor
$\prod_{i}\langle\bar{\psi}_{p_{i}}|\bar{\psi}_{\sigma_{i}(q)}\rangle$
reduces to $x^{j}$. The case of unequal indistinguishability is treated
elsewhere \cite{Renema2018a}. 

\section{Error theory for a Fock state input}

To introduce our techniques for complexity analysis, we review the
derivation of the simulability criterion for the case where the input
state $|\Psi\rangle$ is a product of Fock states. This section is
a summary of \cite{Renema2018a,Renema2018b}, and the supplemental
material included therein. 

For the case of a product-of-Fock-states input, the double sum over
$\xi$ in eqn 6 drops out since there is only a single $\xi=|\Psi\rangle$,
and we are left with:
\begin{equation}
P(\phi_{pd})=\sum_{j=0}^{m}x^{j}\sum_{\sigma^{j}}\sum_{\rho}\mathrm{Perm(M_{1,\rho}\circ M_{\sigma_{p},\rho}^{\dagger})\mathrm{Perm}}(|M_{1,\bar{\rho}}|^{2}).
\end{equation}

We wish to construct an algorithm to approximate $P$ by some efficiently
computable quasiprobability $P'.$ The strategy which we will follow
is to truncate the outer sum of eqn 7 at some value $k<m.$ The intuition
for this is that for all $x<1$, the terms in eqn 7 are exponentially
surpressed by the effect of partial distinguishability. We will show
that, averaged over $M,$ the sum over permanents in eqn 7 is of equal
magnitude for all $j$. This means that $P(\phi_{pd})$ can be interpreted
as a polynomial in $x$ with coefficients of order 1, where the higher
power terms can be neglected when the polynomial is evaluated at sufficiently
small values of $x$. 

The natural way to evaluate the quality of such an approximation is
to compute the variational distance $d=\frac{1}{2}\sum_{\phi}|P(\phi)-P'(\phi)|$,
where the sum runs over all possible output configurations $\phi$
of a boson sampler. The first thing to note is that $d$ is a function
of the matrix $U$ associated with a given boson sampler, which is
highly inconvenient. Therefore, we look instead at the average $E_{U}(d)$,
where the average is taken over the Haar measure of unitary matrices.
Such an average can be related to the value of $d$ of any arbitrary
matrix $U$ by a Markov inequality: the probability that a given matrix
$U$ has a value $d(U)>cE_{U}(d)$ (where $c$ is a positive constant)
is at most $1/c$. If we are willing to accept some small probability
$\delta$ of failure in our algorithm, it therefore suffices to compute
the average of $d$ over all unitaries.

In order to compute $E_{U}(d)$, we make the assumption that the dimension
$N$ of our matrix $U$ is much larger than the number of photons
$m$. In this limit, the correlations between elements of $U$ due
to the unitary constraint can be neglected; the elements approach
independent complex Gaussians with mean $\mu=0$ and standard deviation
$\sigma=1/\sqrt{2N}$ in both real and imaginary part. This limit
is the situation which which the hardness of boson sampling is believed
to hold. Furthermore, in this situation, all output configurations
are equivalent, the average probability of each individual outcome
of the sampler is given by $m!/N^{m},$ and the probability of collision
events (with two or more bosons emerging from the same output port)
can be neglected. For these reasons, it suffices to compute the error
$E(\Delta P)=E(|P(\phi)-P'(\phi)|)$ of a single outcome, and show
that it is of the form $E(\Delta P)=Cm!/N^{m},$ with $C$ some constant
\cite{Renema2018b}. In that case, the average variational distance
is given by $E_{U}(d)=C.$ 

To compute the error $\Delta P$ on a approximating a single output
probability, we define $c_{j}=\sum_{\sigma^{j}}R_{\sigma},$ with
$R_{\sigma}=\Re\left(\sum_{\rho}\mathrm{Perm(M_{1,\rho}\circ M_{\sigma_{p},\rho}^{\dagger})\mathrm{Perm}}(|M_{1,\bar{\rho}}|^{2})\right).$
Hence we can rewrite eqn 7 as: 

\begin{equation}
P(\phi_{pd})=\sum_{j}c_{j}x^{j}=\sum_{j}x^{j}\sum_{\sigma^{j}}R_{\sigma}.
\end{equation}

The first thing to note is that in our definition of $R_{\sigma}$,
it suffices to consider only the real parts of the terms in eqn 7.
This is due to the fact that $R_{\sigma}=R_{\sigma^{-1}}^{\dagger}$,
meaning that imaginary contributions to eqn 8 cancel pairwise. Therefore,
we may concern ourselves only with the real parts of the terms.

Next, we compute the variance of $c_{j}$:

\begin{equation}
\mathrm{var}(c_{j})=\sum_{\sigma}\mathrm{var}(R_{\sigma})+\sum_{\sigma,\tau}\mathrm{cov}(R_{\sigma},R_{\tau}).
\end{equation}

Since $E(R_{\sigma})=0,$ $\mathrm{cov}(R_{\sigma},R_{\tau})=E(R_{\sigma}R_{\tau}).$
This is only nonzero if the parts of $\sigma$ and $\tau$ which carry
phase are each others inverses. The part of $\sigma$ and $\tau$
which carries phase is the entire permutation apart from the fixed
points. We will denote the non-fixed part of a (partial) permutation
$\sigma$ as $\sigma_{p}$, so that the requirement can be stated
as $\sigma_{p}=\tau_{p}^{-1}$. The reason for this is that when averaging
over the Haar measure, each matrix element has an independent, uniformly
distributed phase, which means that that matrix element averages out
to zero. This is also true for products of matrix elements, unless
each phase is canceled pairwise by its conjugate phase, as happens
when $\sigma_{p}=\tau_{p}^{-1}$ \cite{Renema2018a}. Note that while
the fixed points may in principle be different, in the specific scenario
which we are considering now, where there is only one $\xi,$ (i.e.
product of Fock states input) the condition $\sigma_{p}=\tau_{p}^{-1}$
enforces $\sigma=\tau^{-1}.$ Importantly, this will not be true in
the general case which we will treat later on.

Continuing our computation of $\mathrm{var}(c_{j}),$ it turns out
that if $\sigma_{p}=\tau_{p}^{-1}$ , then $1/e<\mathrm{cov}(R_{\sigma},R_{\tau})/\sqrt{\mathrm{var}(R_{\sigma})\mathrm{var}(R_{\tau})}<1.$
Therefore, the problem of estimating $\mathrm{var}(c_{j})$ consists
essentially of estimating $\mathrm{var}(R_{\sigma}),$ and of counting
the number of matches between all possible permutations $\sigma$
and $\tau$ \cite{Renema2018a}. $\mathrm{var}(R_{\sigma})$ is a
function of $j,$ which by definition is the size of $\sigma_{p}$.
For $\mu(\xi)=1,$ it is given by: 
\begin{equation}
\mathrm{var}(R_{\sigma})=e\left(\begin{array}{c}
m\\
j
\end{array}\right)(m-j)!^{2}j!/2N^{2m}.
\end{equation}

Since for the case of a single $\xi$, $\sigma_{p}=\tau_{p}^{-1}$
implies $\sigma=\tau^{-1},$ in this case we have:
\begin{equation}
\mathrm{var}(c_{j})<2\mathrm{var}(R_{\sigma})R(m,m-j).
\end{equation}

where $R(m,m-j)$ is the rencontres number, which counts the number
of permutations of size $m$ with $m-j$ fixed points. In the limit
of large $m,$ $R(m,m-j)=\left(\begin{array}{c}
m\\
j
\end{array}\right)j!/e$, hence $\mathrm{var}(c_{j})<m!^{2}/N^{2m}.$ Applying the inequality
$E(|x|)<\sqrt{\mathrm{var}(x)}$, which holds for any variable $x$
with $E(x)=0,$ we upper bound the deviations of $c_{j}$ from zero
as $E(|c_{j}|)<m!/N^{m}.$ Substituting this back into the truncated
version of eqn 7 and summing the error terms, we find that for a single
outcome, the expected error $E_{U}(\Delta P)=E_{U}(|P-P'|)$ is given
by $\Delta P=m!/N^{m}\sqrt{x^{2(k+1)}/(1-x^{2})}$, which is of the
required form. This implies that for a disinguishability level $x,$
a boson sampler is classically simulable on average with error $E_{U}(d)$
at an approximation level $k$ given by $E_{U}(d)=\sqrt{x^{2(k+1)}/(1-x^{2})}.$
Crucially, in the limit of large $m$, the level of truncation $k$
does not depend on the number of photons $m$ (apart from the trivial
point that $k$ must be smaller than $m$), but only on the level
of imperfections $x$. The case of finite $k$ and $m$ is dealt with
in \cite{Renema2018a}, but does not alter the picture substantially.

The next step is to observe that by truncating the sum over $j$ in
eqn 7 at a fixed $k$, we have produced an approximation which can
be computed efficiently. Approximating permanents of matrices of positive
numbers can be done efficiently \cite{Jerrum2001}, whereas the best
known algorithm for computing permanents of arbitrary matrices scales
as $n2^{n}$ with the matrix size $n$ \cite{Glinn2010,Ryser}. By
truncating eqn 7 at some level $k$, we have therefore achieve polynomial
scaling of our approximate distribtion $P'$ with the number of photons
$m$ \cite{Renema2018a}. 

Having found a distribution of which individual elements can be computed
efficiently, and which is close in variational distance to the output
distribution of an imperfect boson sampler, the next task is to convert
this into a sampling algorithm. We do this by running a Markov Chain
Monte Carlo sampler on our approximate distribution \cite{Neville2017}. 

\section{Error theory for arbitrary input state}

In Section I, we recalled the theory for computing the probability
of an outcome of a boson sampler fed with arbitrary quantum states.
We observed that for a fixed detection outcome, this results in a
double sum over all possible ways in which that quantum state could
have given rise to the observed number of photons. 

In Section II, we observed that the way to compute the effect of noise
on a boson sampler is to count all pairs of permutations $\sigma$
and $\tau$ of the photons which match a given pattern, namely $\sigma_{p}=\tau_{p}^{-1}$.

It will not come as a surprise, then, that the way to compute the
effect of noise on a boson sampler fed with arbitrary input states
is to compute the number of ways in which a double sum over input
states can give rise to permutations satisfying this same rule. 

More formally, if we consider applying the scheme described in Section
II to eqn 6, we have:
\begin{eqnarray}
\mathrm{var}(c_{j}) & = & \sum_{p,q,\sigma}\mathrm{c_{p}c_{q}^{\dagger}var}(R(\xi_{p},\sigma(\xi_{q})))+...\\
 &  & \sum_{p,q,r,s,\sigma,\tau}\mathrm{c_{p}c_{q}^{\dagger}c_{r}c_{s}^{\dagger}cov}(R(\xi_{p},\sigma(\xi_{q}))R(\chi_{r},\tau(\chi_{s}))),\nonumber 
\end{eqnarray}
where $R(\xi_{p},\sigma(\xi_{q}))=\Re(\mathrm{Perm}(M_{p}\circ M_{\sigma(q)}^{\dagger}))$,
and where $\xi_{p},\ \xi_{q},\ \chi_{r}$ and $\chi_{s}$ are all
elements of the set of $\xi,$ and $\sigma$ and $\tau$ are permutations.
The sextuple sum is over all valid assignments of these variables
where at least one variable from $\{\xi_{p},\xi_{q},\sigma\}$ differs
from its counterpart. Following the phase cancellation argument presented
above, the condition for nonzero covariance can now be given as an
elementwise set of conditions on the six variables, which we call
the \emph{pairing rules}:
\begin{eqnarray}
 & \forall i\in\{1...m\}: & \left(\left(\xi_{p,i}=\sigma(\xi_{q})_{i}\right)\wedge\left(\chi_{r,\rho_{j}}=\tau(\chi_{s})_{\rho_{i}}\right)\right)\\
 &  & \vee\left(\left(\xi_{p,i}=\tau(\chi_{s})_{\rho_{i}}\right)\wedge\left(\chi_{r,\rho_{i}}=\sigma(\xi_{q})_{i}\right)\right),\nonumber 
\end{eqnarray}

where $\rho$ is some freely chosen permutation of the indices of
$\chi_{r}$ and $\sigma(\chi_{s})$, and $\xi_{p,i}$ denotes the
$i$-th element of the permutation $\xi_{p}$ (and similarly for the
other variables). Equation 13 is just an elementwise restatement of
the requirement that the unfixed points of the permutation must match
up, while the fixed points are free. In other words, for each pair
of elements of $\xi_{p}$ and $\sigma(\xi_{q}),$ (i.e. for each pair
of $\xi_{p,i}$ and $\sigma(\xi_{q})_{i}$) we must either have that
$\xi_{p,i}=\sigma(\xi_{q})_{i},$ (fixed point) or if this is not
the case, we must have that we can uniquely match up this pair to
a counterpart, i.e. $\xi_{p,i}=\tau(\chi_{r})_{\rho_{i}}$ and $\chi_{r,\rho_{i}}=\sigma(\xi_{q})$.
Note that the requirement that this pairing is unique (no two elements
of $\xi$ can be matched to the same $\chi$), combined with the fact
that the pairing rules are symmetric in $\xi$ and $\chi$ automatically
enforces that $\sigma_{p}$ and $\tau_{p}$ are the same size. 

Furthermore, it should be noted that unlike the case of a single product-of-Fock-states
input, we are now dealing with partial permutations instead of complete
permutations. A partial permutation is a bijection between two subsets
of a given set, whereas a permutation is a bijection on the entire
set. The reason we obtain partial permutations is because $\xi_{p}$
and $\xi_{q}$ (and equivalently $\chi_{r}$ and $\chi_{s}$) do not
necessarily contain the same elements. This corresponds to the fact
that we are considering interference between histories where different
sources produced photons. 

To illustrate the pairing rules, and to illustrate how the introduction
of arbitrary input states changes things compared to Fock state inputs,
an example is in order. For our example, will drill down to a single
term in the sextuple sum of eqn 12. In this example, we consider 5
sources labeled 1 through 5 collectively emitting 3 photons. The set
of allowed $\xi$ is given by: $\{\xi\}=\{(1,2,3),$ $(1,2,4),$ $(1,2,5),$
$(1,3,5),$ $(1,4,5),$ $(2,3,4),\ ...\}$ where the numbers indicate
the presence of a photon in a given mode. We now consider an example
of two choices of $R$ have covariance with each other. For illustration,
we pick a term where $\xi_{p}\neq\xi_{q}$, i.e. a cross term in eqn.
5. An example of a cross term is $\xi_{p}=(1,2,4),$ $\xi_{q}=(2,3,4),$
$\sigma=(231),$ meaning $\sigma(\xi_{p})=(4,2,3)$ and giving rise
to the partial permutation (given in two-line notation for clarity)
$(\xi_{p},\sigma(\xi_{q}))=\left(\begin{array}{ccc}
1 & 2 & 4\\
4 & 2 & 3
\end{array}\right)$. In this case, the partial permutation neglecting fixed points is
$\sigma_{p}=\left(\begin{array}{cc}
1 & 4\\
4 & 3
\end{array}\right).$

When considering which other partial permutations this permutation
will have covariance with, we must find one with matching $\tau_{p}.$
For example, taking $\chi_{r}=(3,4,5),$ $\chi_{s}=(1,4,5)$ and $\tau=(213)$
gives $\left(\begin{array}{ccc}
3 & 4 & 5\\
4 & 1 & 5
\end{array}\right)$ and hence $\tau_{p}=\left(\begin{array}{cc}
3 & 4\\
4 & 1
\end{array}\right),$ which satisfies the covariance rule. 

This example illustrates two imporant points regarding the process
of computing the covariance. First: in the case of a coherent superposition
of input states, there is more freedom in chosing pairs of $R$ which
have covariance, since the two permutations may have different fixed
points and still contribute covariance. Secondly, the pairing rules
in equation 13 do not split into pairwise requirements on either $\xi_{p,q}$
and $\chi_{r,s}$ or on $\sigma$ and $\tau$. Indeed, in the example
we have given above, all of $\xi_{p},$ $\xi_{q}$, $\chi_{r}$ and
$\chi_{s}$ are distinct, and $\sigma\neq\tau^{-1},$ yet we can construct
a situation where $\sigma_{p}=\tau_{p}^{-1}$ holds. The reason the
covariance criterion (eqn 13) doesn't split into a criterion on $\xi_{p,q}$
and $\chi_{r,s}$ or on $\sigma$ and $\tau$ is that for each entry,
the criterion can either be satisfied by a relation between $\xi_{p}$
and $\sigma(\xi_{q})$ or by a relation between $\xi_{p}$ and $\tau(\chi_{s})$
(and the corresponding variables). 

In contrast, since the size of $\sigma_{p}$ and $\tau_{p}$ are equal,
the pairing rules do split into a criterion on the permuted and unpermuted
parts of the permutation. We will use this fact later on in computing
the number of possible covariance terms for specific input states.

A few facts from the Fock state case do carry over. In particular,
the fact that $\sigma_{p}$ and $\tau_{p}$ have the same size means
that when we group terms by number of fixed points, we automatically
group the $R$ by their potential covariance partners. This implies
that $c_{j}$ have zero covariance with each other, and hence that
the strategy of computing $\mathrm{var}(c_{j})$ by considering each
$j$ seperately is still sound, as it was for Fock states.

Finally, we note that unlike the case of a Fock state input, for arbitrary
input states, it does not suffice to show that the coefficients $c_{j}$
are bounded in variance (and hence that the sum over $j$ can be truncated).
The reason for this is that for arbitrary inputs, the truncation does
not immediately produce an efficient approximation algorithm. This
is because the sum over $\xi_{p}$ and $\xi_{q}$, which arises in
the arbitrary state case, and which can contain exponentially many
terms. It is not possible to sample over $\xi_{p}$ and $\xi_{q}$,
since the sum is not convex (not all terms are positive). This means
that in order to approximate an outcome, we must evaluate not just
the sum over $j$, but also the sum over $p$ and $q$; we cannot
sample over those terms. However, we will see that it is sometimes
possible to gather terms in such a way that the overall expression
can be efficiently sampled from.

In summary, when encountering a new input state for a boson sampling
experiment, our tasks are twofold:
\begin{enumerate}
\item Enumerate the set of $\xi_{p}$ and corresponding $c_{p}.$ Compute
the covariance between the set of $R$ and hence the truncation point
$k$ by applying the pairing rules and eqs. 10 and 12.
\item Rewrite eqn 6 to a form that can be computed efficiently at that truncation,
given the particular quantum state in question. 
\end{enumerate}

\section{Superposition Boson Sampling}

Having set up the necessary machinery, we are now ready to investigate
the simulability properties of some optical systems of interest. We
begin with a model which we name \emph{superposition sampling}. In
this model, there are $n$ photon sources, each of which emits the
quantum state $|\psi\rangle=\cos(\alpha)|0\rangle+\sin(\alpha)|1\rangle.$
These sources form a product state at the input of the interferometer:
$|\Psi_{\mathrm{sbs}}\rangle=|\psi\rangle^{\varotimes n}|0\rangle^{\varotimes N-n}$.
Since each source can emit at most one photon, the set of $\xi$ is
given by the ${n \choose m}$ ways of selecting $m$ photons from
$n$ sources. Since all configurations are equally likely and all
multiplicities are equal to 1, we can normalize the probabilities
assuming that exactly $m$ photons are detected, in which case they
are given by $c_{p}=\sqrt{{n \choose m}^{-1}}$ for all $p.$

This problem is motivated by two facts. First, it is of interest in
its own right, due to the fact that it can be implemented using quantum
dot sources \cite{Loredo2018}. Second, it will provide a helpful
stepping stone to the analysis of Gaussian boson sampling further
on. 

To analyse the simulability of this model, we compute $\mathrm{var}(c_{j})$
by counting the number of ways in which the pairing rules are satisfied
for a given $j,$ as a function of $m$, $n$ and $j.$ Since, as
we observed in section III, the paring rules split into a requirment
corresponding to $\sigma_{p}$ and into one on the fixed points, we
can count the number of ways to satisfy the pairing rules by first
assigning the size of $\sigma_{p}$, then the number of ways to create
$\sigma_{p}$, and then the number of fixed points. 

As an illustration of this combinatorics problem and to show how the
paring rules work in practice, we will fill in a table of all variables
to which we must assign a value as we go along with our computation
of the number of possible assignments. An example of a valid assignment
of all variables is given by:
\begin{equation}
\left(\begin{array}{c}
\xi_{p}\\
\sigma(\xi_{q})\\
\chi_{r}\\
\tau(\chi_{s})
\end{array}\right)=\left(\begin{array}{cccccc}
1 & 2 & 3 & 4 & 5 & 6\\
1 & 4 & 2 & 3 & 6 & 5\\
2 & 3 & 4 & 5 & 6 & 7\\
3 & 4 & 2 & 6 & 5 & 7
\end{array}\right),
\end{equation}
which meets the condition that $\sigma_{p}=\tau_{p}^{-1}$. 

We begin by noting that since the set of $\xi$ are symmetric under
permutation of indices, their choice is entirely equivalent. In our
table we can choose $\xi_{p}=\xi_{1}=\{1...m\}$ out of all ${n \choose m}$
possibilities without loss of generality. 
\[
\left(\begin{array}{ccccccc}
1 & 2 & 3 & 4 & 5 & ... & m\\
\emptyset & \emptyset & \emptyset & \emptyset & \emptyset & \emptyset & \emptyset\\
\emptyset & \emptyset & \emptyset & \emptyset & \emptyset & \emptyset & \emptyset\\
\emptyset & \emptyset & \emptyset & \emptyset & \emptyset & \emptyset & \emptyset
\end{array}\right),
\]
where the empty set symbol $\emptyset$ represents an as yet unassigned
value. 

Next, we choose the $m-j$ fixed points of $\sigma,$ i.e. the number
of times we will satisfy the first clause of eqn 13. For $m-j$ fixed
points, this can be done in ${m \choose j}$ ways (chosing the positions
that will not be fixed points), giving a total of ${n \choose m}{m \choose j}$
ways of reaching this step, and an example table of:

\[
\left(\begin{array}{ccccccc}
1 & 2 & 3 & 4 & 5 & ... & n\\
1 & \emptyset & 3 & \emptyset & \emptyset & \emptyset & \emptyset\\
\emptyset & \emptyset & \emptyset & \emptyset & \emptyset & \emptyset & \emptyset\\
\emptyset & \emptyset & \emptyset & \emptyset & \emptyset & \emptyset & \emptyset
\end{array}\right),
\]

Where we have picked some arbitrary assignment of fixed points for
illustration. Next, we must assign the unpermuted part of $\sigma.$
We have already assigned $m-j$ elements, hence we have${n-m+j \choose j}$
ways of chosing the remaining elements, and we have $j!$ ways of
arranging these elements. Note that we overcount here: we have neglected
to exclude those permutations which introduce additional fixed points.
However, in this we overcount by at most a constant factor: the probability
that a full permutation includes a fixed point is $1/e$, and it approaches
zero for highly partial permutations, i.e. when $m\ll n$ . Therefore,
we have ${n \choose m}{m \choose j}{n-m+j \choose j}j!$ ways of reaching
the following table (again with arbitrary numbers for illustration): 

\[
\left(\begin{array}{ccccccc}
1 & 2 & 3 & 4 & 5 & ... & n\\
1 & 4 & 3 & 7 & 6 & ... & 2\\
\emptyset & 4 & \emptyset & 7 & 6 & ... & 2\\
\emptyset & 2 & \emptyset & 4 & 5 & ... & n
\end{array}\right).
\]

Observe that by fixing $\sigma_{p},$ we have fixed the corresponding
elements of $\tau_{p}$ as well. What remains to be done is to pick
the remaining fixed points in $\tau.$ This can be done independently
of the fixed points in $\sigma$ since the fixed points need not match
between $\sigma$ and $\tau$ (as noted above), but not independently
from $\tau_{p},$ since we cannot re-use elements which we have used
there. Note that any choice of fixed points produces a valid $\chi_{r}$
since any choice of $m$ distinct elements from $\{1...n\}$ is a
valid $\chi_{r}$, and by construction our method always arrives at
such a choice. In fixing $\sigma_{p}$, we have used at least $j$
elements, and hence we have to chose the remaining $n-j$ fixed points
from $m-j$ possible choices, hence we have ${n \choose m}{m \choose j}{n-m+j \choose j}{n-j \choose m-j}j!$
ways of filling out the entire table. Therefore, for the case of superposition
sampling, the covariance term in eqn 12 is upper bounded by: 
\begin{eqnarray}
 & \sum_{p,q,r,s,\sigma,\tau}\mathrm{cov}(R(\xi_{p},\sigma(\xi_{q}),R(\chi_{r},\tau(\chi_{s}))\nonumber \\
< & \left(\begin{array}{c}
n\\
m
\end{array}\right)\left(\begin{array}{c}
m\\
j
\end{array}\right)\left(\begin{array}{c}
n-m+j\\
j
\end{array}\right)\left(\begin{array}{c}
n-j\\
m-j
\end{array}\right)j!\mathrm{\mathrm{var}(}R).
\end{eqnarray}

Note that since $\mathrm{cov(R,R')<}\mathrm{var(R)}$, we can simply
ignore the variance term when obtaining an upper bound, if we allow
the covariance term to run over all assignments of $p$, $q$, $r$,
$s$, $\sigma$ and $\tau$.

It is instructive to consider the two extremal cases. If $j=0$ (i.e.
the classical interference term), then $\sigma$ must consist entirely
of fixed points. This can only be achieved by setting $\xi_{q}=\xi_{p}$
and $\sigma=I$, leaving only the ${n \choose m}$ choices of $\xi_{p}$
free. Similarly, we are free to choose $\chi_{r}$, but then the pairing
rules enforce that $\tau=I$ and $\chi_{s}=\chi_{r}$. For this reason,
there are ${n \choose m}^{2}$ covariance terms for $j=0$. In that
case, $\mathrm{var}(c_{j})<m!^{2}/N^{2m},$ since the factor ${n \choose m}^{2}$
drops out against the factor $c_{p}^{4}={n \choose m}^{-2}$ which
arises from the normalization. 

For the other extremal case, we have $j=m.$ In this case, we again
have ${n \choose m}$ ways of chosing $\xi_{p},$ but now all choices
of $\xi_{q}$ and $\sigma$ are good (since we are upper bounding,
we neglect those cases where our choice of $\sigma$ introduces further
fixed points). Therefore, we can pick $\xi_{p},$ $\xi_{q}$ and $\sigma$
without any constraints, and we have ${n \choose m}^{2}m!$ ways of
doing so. However, we now have no freedom left at all in chosing $\tau,$
$\chi_{r}$ and $\chi_{s}$: we must chose $\tau=\sigma^{-1},$ $\xi_{p}=\chi_{s}$
and $\xi_{q}=\chi_{r}.$ Therefore, there are $m!$ more permutations
than in the case $j=0,$ but this drops out against the fact that
these permutations are each a factor $m!$ smaller in value as well,
as given by eqn 10. Hence for $j=m$ we also have $\mathrm{var}(c_{j})=m!^{2}/N^{2m}.$ 

To discuss the simulability of superposition sampling with imperfections,
we first observe that the ratio $\mathrm{var}(c_{j})/\mathrm{var(}c_{0})$
is unbounded. If we take the limit $m\ll n$ and $j\ll m$, then the
variance reduces to $\mathrm{var}(c_{j})={n-j \choose m-j}{n-m+j \choose j}{n \choose m}^{-2}m!^{2}/N^{2m}\approx{m \choose j}m!^{2}/N^{2m},$
which becomes arbitrarily large around $j=m/2$ when $m$ is increased,
and where the last equality holds at large $n.$ This leaves open
the possibility that superposition sampling is tolerant to imperfections. 

However, numerical simulations suggest this is not the case: when
converting from $\mathrm{var}(c_{j})$ to $E(|c_{j}|),$ the inequality
$\sqrt{\mathrm{var}(c_{j})}>E(|c_{j}|)$ becomes less and less tight
as $n$ grows, resulting in coefficients $\mathrm{var(c_{j})\approx}\mathrm{var(}c_{0})$
for all $j.$ We have observed this effect in simulations from $m=2$
to $m=6,$ and from $n=2$ to $n=25,$ (but not all combinations of
both) averaging over 1000 Haar-random matrices in each case. In those
same simulations, we also computed $\mathrm{var}(c_{j})$ for each
$n,$ $m,$ and $j$, and we observed that (accounting for some finite
size effects which give small corrections), our computations of $\mathrm{var}(c_{j})$
as described above match our simulations. For large $n,$ we find
$\mathrm{var}(c_{j})>\mathrm{var}(c_{0}),$ as suggested by eqn 15.
This shows that the limitations of our method are really in the conversion
from variance to absolute moment. We leave this aspect of the problem
for future study. 

Next, we must rewrite eqn 6 to a form which can be efficiently sampled
from. In the form in which it is given, there are exponentially many
terms in the sum over $\xi_{p}$ and $\xi_{q}$, which means that
a truncation at $j$ will not lead to efficient sampling. This can
be remedied by the following observation: for a given $\sigma_{p},$
we know that we will encounter every possible assignment of the ${n \choose m-j}$
remaining possible fixed points, and with equal weight (since all
$c_{p}$ are equal). However, each assignment occurs in a different
pair of $\xi_{p}$ and $\xi_{q}$. Grouping terms, we have:
\begin{eqnarray}
P & = & \sum_{j=0}^{m}\sum_{\sigma_{p}^{j}}\sum_{\rho}\mathrm{Perm}(M_{I_{\sigma_{p}},\rho}\circ M_{\sigma_{p},\rho}^{\dagger})\times...\nonumber \\
 &  & \mathrm{Perm}(M_{\mathrm{fp}}(\sigma_{p},\rho))/(n-m-j)!,
\end{eqnarray}
where $I_{\sigma_{p}}$ denotes the elements of $\sigma_{p}$ in sorted
order (i.e the partial identity permutation of the elements of $\sigma_{p}$),
where $\rho$ is a $j$-partition of $m$, and where $\sigma_{p}^{j}$
denotes all possible partial permutations of size $j$ without fixed
points. The matrix $M_{\mathrm{fp}}(\sigma_{p},\rho)$ is an $n-2j$
by $n-2j$ matrix constructed by the following method: first, construct
an $m$ by $n$ matrix of all norms squared, i.e. $M_{\mathrm{fp},ij}=|M_{ij}|^{2},$
where $M$ is defined according to Section I. Then, delete the $j$
rows corresponding to $\rho,$ and the $2j$ columns corresponding
to $\sigma_{j}^{p}$. A matrix of size $m-j$ by $n-2j$ remains.
Next, pad out this matrix with $n-m-j$ rows of ones to make a square
matrix. Laplace expansion along these rows shows that this produces
precisely every way of picking fixed points that is complementary
with a given choice of $\sigma_{j}^{p}$ and $\rho,$ and the factor
$1/(n-m-j)!$ compensates for the spurious permanents of ones introduced
by this procedure. 

Sampling from eqn 16 can be done efficiently: since there are ${n \choose 2j}$
ways of choosing $\sigma_{p}$, when truncating at a fixed $k=j,$
this goes approximately as $n^{2k}$. Furthermore, the permanent of
$M_{\mathrm{fp}}$ is a permanent of a positive matrix of size $n-j,$
so it can be evaluated at a cost polynomial in $n.$ Since the typical
expected number of photons is proportional to $n$ via $\bar{m}=\sin^{2}(\alpha)n,$
this means that the scheme is polynomial in $m.$ 

Building on our numerical observations regarding the behaviour of
$E(|c_{j}|)$, we can compute the error level of trunctation of the
outer sum of eqn 16 at a given $k$. This computation can be reduced
to a known case if we note that we did not move any terms between
different $j$ when going from eqn 6 to eqn 16. Introducing a distinguishability
factor $x^{j}$, truncating and summing the series, we find that superposition
boson sampling is as loss tolerant as regular boson sampling, with
an expected error of $E(d)=\sqrt{x^{2(k+1)}/(1-x^{2})}$, as reported
previously for regular boson sampling. 

We observe an interesting parallel between superposition boson sampling
and Fock state boson sampling with loss: in lossy boson sampling,
the input density matrix is of the form $\rho=\left(|0\rangle\langle0|+|1\rangle\langle1|\right)^{\varotimes n}\left(|0\rangle\langle0|^{\varotimes N-n}\right)$.
In that case, there is a similar sum over ${n \choose m}$ sources
as in superposition boson sampling, representing the ${n \choose m}$
ways in which the $n$ sources could have given rise to the $m$ observed
photons. However, in this case it is a single sum, since the summation
sum is incoherent (i.e. there are no interference terms). When comparing
this case to superposition boson sampling, this amounts to considering
only terms $\xi_{p}=\xi_{q}$. Interestingly enough, lossy Fock state
boson sampling is known to be classically simulable because the $c_{j}$
decrease exponentially with $j$ \cite{Renema2018b}. Therefore, the
following physical picture arises: if we think of superposition boson
sampling as a combination of loss terms (that is, terms which would
show up in lossy Fock state boson sampling, where $\xi_{p}=\xi_{q}$)
and cross terms (where $\xi_{p}\neq\xi_{q})$, then the loss terms
are skewed to low photon number interference, while the cross terms
are skewed to high photon number interference, thereby precisely cancelling
out the deleterious effects of the loss terms.

We conclude this section with a few remarks on the viability of superposition
boson sampling as a quantum advantage demonstration. The observation
that the coefficients of perfect superposition boson sampling do not
decay to zero shows that it is not simulable by our method in the
ideal case, and that the state $|\Psi_{\mathrm{sbs}}\rangle$ might
therefore be a valid state to use for a demonstration of a quantum
advantage. However, the fact that this state is no more loss tolerant
than a regular Fock state means that there is no obvious reason to
use this state to achieve a quantum advantage. Rather, the interest
lies in quantum simulation: it is an open problem which physical systems
can be simulated with photonics, and any new imput state which can
be shown to have computationally interesting properties is a potential
addition to our simulation arsenal. The potential arrival of a new
class of states to sample from is valuable, since this could broaden
the range of problems which can be simulated in photonics. The question
whether there are natural problems which can be simulated using superposition
states is one of high interest.

We stress that these results do not constitute a hardness proof, however
the fact that this state passes the most stringent classical simulation
criteria known is some cause for optimism. It is hoped that these
results will spur interest in a hardness proof for superposition sampling. 

\section{Gaussian Boson Sampling}

Finally, we turn our attention to Gaussian boson sampling (GBS). In
GBS, the input quantum state is no longer factorizable across modes,
but is instead factorizable over pairs of optical modes: $|\psi\rangle=\cosh(r)^{-1}\sum_{j=0}^{\infty}(-e^{i\phi}\tanh(r))^{j}|j,j\rangle,$
and $|\Psi\rangle=|\psi\rangle^{\varotimes n}|0\rangle^{\varotimes N-n}$,
where $r$ is a parameter that measures the strength of the optical
squeezing, which we shall assume to be equal for all sources. Again,
we have two tasks: first, to construct the set of $\xi_{p}$ and $c_{p}$,
and apply the pairing rules and eqs 10 and 12 to determine the simulability
criterion, and secondly to rewrite eqn 6 to enable efficient sampling. 

We will do these tasks in order. The strategy will be to build on
the results from SBS. We will show that in the limit of weak squeezing,
GBS reduces to SBS with a single additional constraint. This will
enable us to re-use many of the results from SBS.

Without loss of generality, we exclusively consider two-mode squeezed
vacuum states here. There also exist single-mode squeezed vacuum states,
in which photon pairs are emitted into a single mode. In boson sampling,
the two cases are equivalent, as a layer of 50/50 beam splitters converts
pairs of single mode squeezed states into two-mode squeezed states,
and vice versa. If we think of $U$ as made up of a layer of 50/50
beam splitters and a remaining matrix $U'$, by the definition of
the Haar measure, the probability density of drawing $U$ is as large
as drawing $U'$. Hence every boson sampling experiment with single-mode
squeezers can be thought of as equivalent to an equiprobable boson
sampling experiment with two-mode squeezers. This shows that the two
are equivalent. 

We begin by determining the set of $\xi.$ We again consider the case
of $n$ modes containing photons and $m$ detected photons. Compared
to SBS, there is an additional constraint on $\xi$, which is due
to the bipartite nature of the state: when choosing a photon from
a particular mode, we must also chose a photon from the conjugate
mode. We shall refer to such a pair of photons as a biphoton. Secondly,
the multipair terms in $|\psi\rangle$ mean we also have an additional
options in picking $\xi,$ because we may now pick more than one photon
pair from a given pair of modes. Therefore, the set of $\xi_{p}$
contains ${n/2-m/2-1 \choose m/2}$ elements, reflecting the fact
that we are picking $m/2$ biphotons with replacement. To determine
the $c_{p},$ we note that due to the exponential prefactors in $|\psi\rangle,$
all $\xi_{p}$ occur with the same probability amplitude, namely $\tanh(r)^{m/2}$;
the phases may be absorbed into the action of the interferometer.
However, their multiplicities are no longer equal: $c_{p}=\tanh(r)^{m/2}/\sqrt{\mu(\xi_{p})}.$

To simplify our calculations, we will focus on the case of weak squeezing.
If $(m/2)^{2}\ll n/2,$ then the fraction of $\xi_{p}$ terms where
we pick two biphotons from the same source can be made arbitrarily
small. In that case, we can ignore the arbitrarily small fraction
with nonunit multiplicity, and renormalize the $c_{p}$ to postselect
on a particular photon number outcome (as we did for SBS), giving
$c_{p}=\sqrt{{n/2 \choose m/2}}$. Hence in the limit of low squeezing,
there are ${n/2 \choose m/2}$ choices for $\xi,$ all equiprobable. 

Note that the above shows that in the limit of sufficiently weak squeezing,
GBS strongly resembles SBS. The only difference is the additional
constraint of picking biphotons. This fact will enable us to use the
results from SBS with minor modifications.

We again seek to compute $\mathrm{var}(c_{j})$ by looking for pairs
of terms with matching permutations. For $j=0$ and $j=m$, our results
from SBS carry over in a straightforward fashion. For $j=0,$ we again
have complete freedom in chosing $\xi_{p}$ and $\chi_{r},$ but the
remaining variables are fixed by the pairing rules ($\sigma=\tau=I$
, $\xi_{p}=\xi_{q}$, $\chi_{s}=\chi_{r}$). Therefore, we have ${n/2 \choose m/2}^{2}$
ways of creating matching $R,$ and $\mathrm{var}(c_{j})=m!^{2}/N^{2m},$
as before. Similarly for $j=m,$ all $\xi_{p}$, $\xi_{q}$ and $\sigma$
are allowed, but their combination completely fixes $\tau,$ $\chi_{r}$
and $\chi_{s}$. Hence we have $\mathrm{var}(c_{m})<m!^{2}/N^{2m}$
as before. 

For the case of arbitrary $j$, the situation is a little more complex.
The biphoton nature of the light field introduces additional constraints.
In particular, if we have biphotons which are not wholly included
in either the fixed points or in $\sigma_{p},$ then these reduce
the number of possible covariance pairings, since the presence of
exactly one photon of a pair in $\sigma_{p}$ implies the presence
of its partner photon in the fixed points (since we must respect the
biphoton pairing). Moreover, the presence of an 'incomplete' biphoton
in $\sigma_{p}$ implies that the same biphoton must occur in $\tau_{p}$
as well, thereby partially determining the fixed points of $\tau$
as well.

This is best illustrated with an example. Consider the case of $j=2$,
for large $m$ and $n$. There are two options for constructing $\sigma_{p}$:
either $\sigma_{p}$ contains two complete biphotons, one in each
row, for example, $\sigma_{p}={1\ 2 \choose 3\ 4},$ or it contains
two incomplete biphotons, for example $\sigma_{p}={1\ 3 \choose 3\ 1}$.
Note that these are the only options allowed if we take into account
the fact that the remaining elements of $\sigma$ must be fixed points.
For example, if we had $\sigma_{p}={1\ 3 \choose 2\ 1}$, then we
cannot complete the permutation using only fixed points, since the
presence of a $1$ in $\xi_{p}$ implies that mode $2$ must be selected
as well, but since mode $2$ already enters in $\sigma(\xi_{q})$,
it cannot also form a fixed point. Furthermore, for the case of $\sigma_{p}={1\ 3 \choose 3\ 1}$,
we know that both $\tau$ and $\sigma$ must include the fixed points
2 and 4. This example illustrates that the constraint placed on $\sigma$
and $\tau$ is that biphotons which are not completely in $\sigma_{p}$
(or $\tau_{p}$) must occur both in $\sigma$ and $\tau$. 

Since biphotons which straddle $\sigma_{p}$ and the fixed points
determine a fixed point in both $\sigma$ and $\tau,$ while biphotons
contained in the fixed points can be independently chosen between
$\sigma$ and $\tau,$ the leading contribution to the covariance
(expressed in powers of $n$ and $m$) in eqn. 12 at large $m$ and
$n$ will be the one with the fewest straddling fixed points. Therefore,
for even $j,$ it is the one where no fixed points are straddling
$\sigma_{p}$ and the fixed points, and for odd $j,$ there will be
only one. This removes the constraints between fixed points and $\sigma_{p}$
introduced by the biphotons, meaning that we can follow the argument
from SBS to count the number of variance pairs, substituting $n\rightarrow n/2,$
$m\rightarrow m/2,$ $j\rightarrow\lceil j/2\rceil$ in all terms
related to filling the fixed points and $\sigma_{p}$. This means
that the number of terms with covariance is: $\#R=j!{n/2 \choose m/2}{m/2 \choose \lceil j/2\rceil}{n/2-m/2+\lceil j/2\rceil \choose \lceil j/2\rceil}{n/2-\lceil j/2\rceil \choose m/2-\lceil j/2\rceil},$
leading to a covariance of $\mathrm{cov(}c_{j})=\#R\mathrm{var}(R)={n \choose m}^{-1}{m/2 \choose \lceil j/2\rceil}{n/2-m/2+\lceil j/2\rceil \choose \lceil j/2\rceil}{n/2-\lceil j/2\rceil \choose m/2-\lceil j/2\rceil}{m \choose j}$$(m-j)!^{2}$\\
$j!^{2}/N^{2m}$$<m!^{2}/N^{2m}$ for all $j,$ using the inequality
${m \choose j}>{m/2 \choose j/2}^{2}$. 

From this we conclude that like superposition boson sampling, the
higher order interference terms of Gaussian boson sampling are at
least as sensitive to photon distinguishability as in regular boson
sampling. Note that unlike superposition boson sampling this derivation
does not rely on numerical results. When we numerically evaluate eqn.
12 for Gaussian boson sampling, we observe a sawtooth pattern in the
$c_{j},$ where odd $j$ are strongly surpressed. This is a reflection
of the biphoton nature of the light, and it suggests that further
analysis might develop a tighter bound. 

Finally, we turn to the issue of rewriting eqn 6. into a form which
can be efficiently sampled from. We will see that it is possible to
regroup terms to gather identical quantum interference terms, as we
did for the case of SBS. Furthermore, this results in polynomially
many quantum interference terms of fixed size, which means they can
be efficiently computed. Surprisingly, it is the classical interference
terms which are problematic: we obtain an expression for the classical
interference terms for which no known classical algorithm exists to
compute the output probability. 

We begin by following the same strategy as with SBS, by grouping terms
which correspond to the same quantum interference process (i.e. the
same permuted part of the permutation $\sigma_{p}$), but which have
different fixed points. This can be done straightforwardly:
\begin{equation}
P=\sum_{j=0}^{m}\sum_{\sigma_{p}^{j}}\sum_{\rho}\mathrm{Perm}(M_{I_{\sigma_{p}},\rho}\circ M_{\sigma_{p},\rho}^{\dagger})\times\sum_{\bar{\sigma_{p}^{j}}}\text{\ensuremath{\mathrm{Perm}}(}|M_{\sigma_{p}^{j}}|^{2}),
\end{equation}
where $\bar{\sigma_{p}^{j}}$ is the complement of $\sigma_{p}^{j},$
in the sense that it contains a way of chosing fixed points such that
the combination of $\sigma_{p}^{j}$ and $\bar{\sigma_{p}^{j}}$ forms
a 'legal' permutation, i.e. one that corresponds to a particular choice
of $\xi_{p},$ $\xi_{q}$ and $\sigma.$ Note that there are multiple
ways of completing a permutation from its unfixed points, hence there
is a sum over the various possible $\bar{\sigma_{p}^{j}}$ in eqn
17.

To assess the complexity of sampling from eqn 17, many of the arguments
from eqn 16 carry over: there are ${m \choose 2j}$ ways of chosing
$\sigma_{p}^{j}$, hence there are polynomially many quantum interference
terms of size at most $j.$ However, the classical interference term
$\sum_{\bar{\sigma_{p}^{j}}}\text{\ensuremath{\mathrm{Perm}}(}|M_{\sigma_{p}^{j}}|^{2})$
cannot be computed efficiently. This can be seen already when considering
$j=0:$ in this case, there are ${n-m/2+1 \choose m/2}$ ways of picking
sets of $m$ fixed points, which grows exponentially with $m.$ 

As we have done with eqn 16, we will try to recast each classical
interference term in eqn 17 as a permanent of a positive matrix. We
will not succeed because of the biphoton nature of the light field,
which places additional constraints on $\bar{\sigma_{p}^{j}}$. This
is in contrast to SBS, where every choice of fixed points was allowed. 

For biphotons which belong half to a fixed point, half to $\sigma_{p},$
a straightforward construction similar to that described for SBS can
be followed. For the fixed points where both parts of the biphoton
belong to the fixed points, a more elaborate construction is necessary,
which we will show for the case $j=0.$ Define an $m+n$ by $m+n$
matrix $M_{\mathrm{gfp}}:$

\begin{equation}
M_{\mathrm{gfp}}=\left(\begin{array}{cc}
\bar{M} & 0\\
P & S
\end{array}\right)
\end{equation}

with $\bar{M}$ an $m$ by $n$ matrix containing the mod-squared
values of the elements of $M.$ $P$ is an $n$ by $n$ matrix of
the following form:

\begin{equation}
P=\left(\begin{array}{ccccccccc}
1 & 1 & 0 & 0 & \cdots & 0 & 0 & 0 & 0\\
1 & 1 & 0 & 0 & \cdots & 0 & 0 & 0 & 0\\
0 & 0 & 1 & 1 & \cdots & 0 & 0 & 0 & 0\\
0 & 0 & 1 & 1 & \cdots & 0 & 0 & 0 & 0\\
\vdots & \vdots & \vdots & \vdots & \ddots & \vdots & \vdots & \vdots & \vdots\\
0 & 0 & 0 & 0 & \cdots & 1 & 1 & 0 & 0\\
0 & 0 & 0 & 0 & \cdots & 1 & 1 & 0 & 0\\
0 & 0 & 0 & 0 & \cdots & 0 & 0 & 1 & 1\\
0 & 0 & 0 & 0 & \cdots & 0 & 0 & 1 & 1
\end{array}\right),
\end{equation}
 and $S$ and $n$ by $m$ matrix of the following form:
\begin{equation}
S=\left(\begin{array}{ccccc}
1 & 1 & \cdots & 1 & 1\\
-1 & -1 & \cdots & -1 & -1\\
\vdots & \vdots & \vdots & \vdots & \vdots\\
1 & 1 & \cdots & 1 & 1\\
-1 & -1 & \cdots & -1 & -1
\end{array}\right).
\end{equation}

And the remaining quadrant of $M_{\mathrm{gfp}}$ is composed of zeroes.
We can show that $\mathrm{Perm}(M_{\mathrm{gfp}})/2^{m}m!=\sum_{\bar{\sigma_{p}^{j}}}\text{\ensuremath{\mathrm{Perm}}(}|M_{\sigma_{p}^{j}}|^{2})$
by recursively Laplace-expanding along the $i$-th and $i+1$-st rows
of $P$ and $S$ simultaneously, for odd $i$. If we do this, the
only 2-minors which have a nonzero prefactor are either those where
we pick columns $i$ and $i+1$, which corresponds to deleting a pair
of columns from $\bar{M},$ or those where we pick any pair of columns
from $S,$ which only deletes zeroes in the upper $m$ rows. Either
of these picks up a factor $2$, since $\mathrm{Perm}\left(\begin{array}{cc}
1 & 1\\
1 & 1
\end{array}\right)=\mathrm{Perm}\left(\begin{array}{cc}
1 & 1\\
-1 & -1
\end{array}\right)=2.$ Any choice which mixes $P$ and $S$ has a zero prefactor, since
$\mathrm{Perm}\left(\begin{array}{cc}
1 & 1\\
1 & -1
\end{array}\right)=0.$ Therefore, we can see that this construction respects the pairing
of modes which arises from the biphoton nature of the light field:
either it deletes a pair of photons from consideration, or it 'skips
a turn' and deletes no photons. This ensures that when the auxilliary
matrices $P$ and $S$ are exhausted via repeated pairwise Laplace
expansion, only pairs of terms in $\bar{M}$ which correspond to the
same biphoton remain, which completes the proof. 

The reason why this expression for $M_{\mathrm{gfp}}$ does not result
in an efficient algorithm is that in order to cast the classical interference
as the permanent of a single matrix $M_{\mathrm{gfp}}$, we needed
to introduce negative numbers in that matrix to make this construction
work. This means that we cannot apply the approximation algorithm
of Jerrum, Sinclair and Vigoda \cite{Jerrum2001} to efficiently approximate
the permanent of $M_{\mathrm{gfp}}$, since that algorithm requires
a matrix with only positive elements. We leave the question of whether
such an algorithm exists as an open problem for the theory of matrix
permanents. 

On physical grounds, however, it is entirely possible that such an
algorithm exists, since the probability which we are interested in
corresponds to classical transmission, and this also holds for all
cases which arise when $j\neq0,$ which can be accounted for with
a construction analogous to the one presented above. We note that
nothing in the hardness proof of the original Fock state boson sampling
proposal hinges on the existence of the JSV algorithm. However, as
we noted above, at finite levels of imperfection, all quantum interference
effects are accounted for by the quantum interference terms, which
are of size $k$ or smaller. It would be strange if the hardness of
simulating boson sampling, which we think of as an intrinsically quantum
phenomenon, depends on the hardness of computing a classical transmission
probability.

Another way to solve the difficulty of the classical interference
terms would be to construct an algorithm that directly samples from
an approximation of the state taking into account the effects of noise,
instead of approximating an output probability and feeding that to
an MCMC sampler. This was recently achieved for Fock state boson sampling,
\cite{Moylett2019}, although in the resulting algorithm, the truncation
point $k$ is a rising function of $m,$ meaning that the algorithm
is not efficient. Since $\mathrm{Perm}(M_{\mathrm{gfp}})$ represents
a convex sum over positive terms, such a sampling algorithm would
remove the need to compute the entire sum. Instead, a sample from
the classical interference term would suffice, which can be produced
readily by selecting one of the terms and simulating sending photons
through the interferometer one by one \cite{Aaronson2011}.

Finally, we note that our noise model for Gaussian boson sampling
does not capture all sources of imperfections. There are two effects
which we did not consider, and which we would like to highlight. First,
we considered a limit in which double pair emission is negligible,
i.e. in which all multiplicities are equal to one. Since double pairs
reduce the complexity of the sampling problem by introducing repeated
rows in the permanent, they constitute an imperfection. The second
imperfection which we did not consider is an effect which we name\emph{
layer mixing. }This occcurs when sampling from any input state with
an indeterminate photon number, in the presence of loss. At the beginning
of our derivation, we assumed that we knew the number of photons generated
by the sources. In the presence of loss, this knowledge is imperfect;
there is mixedness between the process where $m$ photons are generated
and none are lost, $m+1$ and 1 is lost, and so on. This effect, which
is not present for Fock state sampling, should not be confused with
the quantum interference between photons originating from different
sources. The presence of layer mixing suggests that perhaps both GBS
and SBS are more susceptible to loss than regular Fock state boson
sampling. We leave these issues to future study. 

However, since all previous examples of error analysis in boson sampling
have shown that errors compound, we speculate that the addition of
these errors will not reduce, but rather increase, the simulability
of Gaussian boson sampling under imperfections. This means that until
further results on this topic arise, it is not unreasonable to assume
that our bound will also hold outside of the weak-pumping regime for
which it was derived.

\section{Conclusions}

In this section, we briefly reprise the results of this work. First,
we have demonstrated how to extend the analysis of the effect of imperfections
on boson sampling to the case of arbitrary input states. We used this
method to show that for two particular choices of input states, namely
Gaussian states and superposition states, the output probability without
noise cannot be efficiently approximated by our methods. This is a
necessary but not sufficient condition for computational hardness
of sampling problems using these states. For superposition sampling,
this is the first indication that this problem is of interest for
sampling.

Then, we introduced the effects of noise. Using a combination of analytic
results and numerical simulations, we showed that the efffect of noise
on both these sampling protocols is identical to that of regular boson
sampling, at least for the particular type of noise (distinguishability)
under consideration here; they are no more or less resillient to this
kind of noise than Fock state boson sampling is. For superposition
states, this leads immediately to a classical algorithm which can
efficiently simulate imperfect superposition boson sampling. For Gaussian
states, the existence of such an algorithm depends on a not-implausible
conjecture in the theory of matrix permanents.

As a final warning not to take our results as a hardness proof for
either Gaussian Boson Sampling or Superposition Boson Sampling, we
give an explicit example \footnote{This example is due to R. Garcia-Patron.}
where there is no computational complexity, but where our simulation
strategy does not work. This is the case where weak coherent states
are incident on the interferometer. In this case, phases and amplitudes
can be propagated efficiently through the interferometer, no entanglement
builds up, and we can sample efficiently from the output distribution.
However, if we follow the analysis outlined above, we would find that
perfect 'coherent state sampling' is not susceptible to out algorithm.
In this case, the very first step in our approach, namely the projection
onto Fock states, destroys the structure (namely the coherent state
amplitudes and phases) which enables efficient classical simulation.
This emphasizes the point that failure of any one simulation strategy
is no guarantee that all simulation strategies will fail. 

Finally, we address some open problems. There are two major open problems
raised by this work. The first is to find a better way to compute
$E(|c_{j}|),$ which does not depend on the inequality $\sqrt{\mathrm{var(c_{j})}}>E(|c_{j}|),$
which is insufficiently tight for our purposes. The second open problem
is how to compute the permanent of $M_{\mathrm{gfp}}$ efficiently. 
\begin{acknowledgments}
This research was supported by NWO Veni. I thank Carlos Anton Solanas
and Pascale Senellart for bringing to my attention the problem of
superposition boson sampling and for discussions. I thank Raul Garcia-Patron,
Valery Shchesnovich, Nicolas Quesada, Helen Chrzanowski, Hui Wang,
Chao-Yang Lu, Reinier van der Meer, and Pepijn Pinkse for discussions. 
\end{acknowledgments}

\end{document}